\documentclass{article}
\pdfoutput=1
\usepackage{spconf,amsmath,graphicx}
\usepackage{amsfonts}
\usepackage{amssymb,amsthm,mathtools,nccmath}
\usepackage[numbers, sort&compress]{natbib}

\usepackage{cuted}
\usepackage{MnSymbol}
\title{blind estimation of room acoustic parameters from speech signals based on extended model of room impulse response}
\name{Lijun Wang, Suradej Duangpummet, Masashi Unoki \thanks{This work was supported by SCOPE Program of Ministry of Internal Affairs and Communications (Grant Number: 201605002) and JSPS-NSFC Bilateral
Joint Research Projects/Seminars (JSJSBP120197416).}}
\address{School of Information Science, Japan Advanced Institute of Science and Technology\\
	 1-1 Asashidai, Nomi, Ishikawa, 923-1292, Japan\\
	 {\{lijun.wang, suradej, unoki\}@jaist.ac.jp}}
\begin{document}
\maketitle
\begin{abstract}
The speech transmission index (STI) and room acoustic parameters (RAPs), which are derived from a room impulse response (RIR), such as reverberation time and early decay time, are essential to assess speech transmission and to predict the listening difficulty in a sound field. Since it is difficult to measure RIR in daily occupied spaces, simultaneous blind estimation of STI and RAPs must be resolved as it is an imperative and challenging issue. This paper proposes a deterministic method for blindly estimating STI and five RAPs on the basis of an RIR stochastic model that approximates an unknown RIR. The proposed method formulates a temporal power envelope of a reverberant speech signal to obtain the optimal parameters for the RIR model. Simulations were conducted to evaluate STI and RAPs from observed reverberant speech signals. The root-mean-square errors between the estimated and ground-truth results were used to comparatively evaluate the proposed method with the previous method. The results showed that the proposed method can estimate STI and RAPs effectively without any training.
\end{abstract}
\vspace{-0.5em}
\section{Introduction}
\vspace{-0.5em}
The quality of sound and intelligibility of speech transmitted in a room 
should be evaluated in order to understand room acoustic characteristics and diagnose the degradation in sound quality. Instead of expensive and laborious listening experiments, objective indices and room acoustic parameters (RAPs), i.e., physical descriptions of room acoustics, can assess sound quality and speech intelligibility. 

Several objective indices and RAPs have been investigated and standardized \cite{kuttruff2016room,MTF,IEC60268,ISO3382}. In IEC 60268-16:2020, a speech transmission index (STI) based on the modulation transfer function (MTF) was used to predict the speech intelligibility of a sound field \cite{MTF,IEC60268}. The essential RAPs and their measurements have also been specified in ISO 3382-1:2009, including reverberation time ($T_{60}$), early decay time (EDT), clarity (early-to-late-arriving sound energy ratio: $C_{80}$ / $C_{50}$), Deutlichkeit (early-to-total sound energy ratio: $D_{50}$), and center time ($T_s$) \cite{ISO3382}. $T_{60}$ is the most essential parameter for representing room acoustic characteristics. 
STI and RAPs can be derived from the room impulse response (RIR), which needs to be measured. However, it is difficult to measure RIR in spaces where the people cannot be excluded. Hence, obtaining STI and RAPs from an observed signal, called blind estimation, is necessary.
\vspace{-0.85em}
\begin{figure*}[t]
 	\label{fig:blockDiagram}
 	\centering
 	 \centerline{\includegraphics[width=1.0\textwidth]{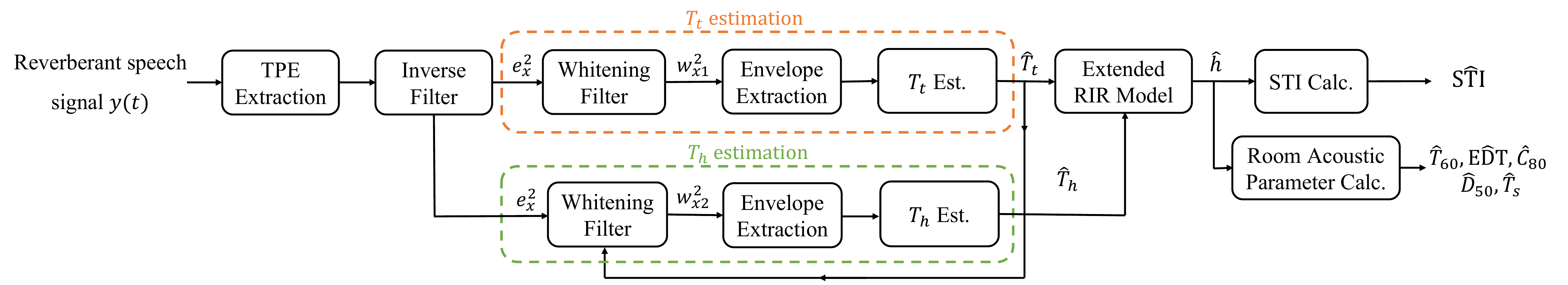}}
 	\caption{Block diagram of proposed method.}
 \end{figure*}
\vspace{-0.5em}
\section{Related works}
\vspace{-0.5em}
Blind estimation methods for STI and RAP have been proposed by using either analytical or learning-based approaches  \cite{Unoki, Unoki2008,Couvreur01model-basedblind, Keshavarz,Ratnam, Kendrick,santos2016blind,Seetharaman2018,LI2007145, Kendrick2008,Callens2020,Steinmetz2021,Suradej}. The analytical approach realizes blind estimation by creating an explicit mapping between observed reverberant signals and desired parameters. M. Unoki {\it{et al.}} proposed two schemes based on the concept of MTF for estimating STI and $T_{60}$  \cite{Unoki, Unoki2008}. The model of speech sequences proposed in \cite{Couvreur01model-basedblind, Keshavarz} and the model of the energy decay curve proposed in \cite{Ratnam, Kendrick}, combined with the maximum-likelihood estimator, were proposed to estimate $T_{60}$. For learning-based methods, many artificial neural networks have been utilized to estimate a desired parameter (e.g., $T_{60}$, $C_{80}$, or STI), including long short-term memory, convolution neural networks (CNN), and recurrent CNN \cite{santos2016blind,Seetharaman2018,LI2007145, Kendrick2008,Callens2020}. Recently, Suradej {\it{et al.}} proposed the MTF-based CNN scheme to estimate STI and RAPs \cite{Suradej}.

However, the current methods can estimate only a single parameter \cite{Unoki, Unoki2008, Couvreur01model-basedblind, Keshavarz, Ratnam, Kendrick}. Although the MTF-based CNN method could estimate multiple parameters, it is limited to the training data used to derive the model, the same as the other learning-based methods \cite{LI2007145, Kendrick2008, santos2016blind, Seetharaman2018, Callens2020, Steinmetz2021, Suradej}. The efficiency of the trained models naturally decreases when the real environments differ from the training data. The models are also difficult to optimize due to them being untraceable implicit models and having a vast number of trainable parameters.
Therefore, we propose an analytical method for blindly estimating the STI and five RAPs, $T_{60}$, EDT, $C_{80}$, $D_{50}$, and $T_{s}$, simultaneously. We incorporate a stochastic RIR model, namely an extended RIR model, into
 the relationships between the temporal power envelope (TPE) of an observed signal and the RIR model to derive the method.
\section{Proposed method}
\label{sec:relwork}
\vspace{-0.5em}
We propose a blind estimation method, the alternating estimation strategy (AES), shown as a block diagram in Fig.\,\ref{fig:blockDiagram}. The details are given as follows.
\vspace{-1.0em}
\subsection{Extended RIR model}
When blindly estimating STI and RAPs, we can observe reverberant signals only.
Thus, we model an observed signal in a reverberant room as the convolution of an original signal and RIR. 
Schroeder's RIR model is a simple decay model and is commonly used to approximate a measured RIR that is unknown \cite{schroeder1981}.
However, it has a limitation because it lacks modeling of the onset transition. 

Figure\,2 shows a comparison between the fits of the temporal power envelope of the RIR models with the measured RIR. Hence, in realistic spaces, Schroeder's RIR model mismatches the actual RIR. As a result, a more accurate RIR model has been proposed, namely the extended RIR model \cite{unoki2016study}. The extended RIR model is defined as:
\begin{equation}
h(t)=e_h(t)\textbf{c}(t)=
\bigg\{\begin{array}{l}
		a \exp(6.9t/T_h)\textbf{c}(t),\hspace{15pt} t<0 \\
		a \exp(-6.9t/T_t)\textbf{c}(t),\hspace{10pt} t\geq0 
    \end{array}
	\label{eq:ext_RIR}
\end{equation}
\begin{equation}
	h_{\rm{ext}}(t) = h(t-t_0), \quad t_0\geq0
	\label{eq:ext_RIR_time-shifting}
\end{equation}
where $T_h$ and $T_t$ denote the parameters controlling the rising and decaying envelopes of the RIR. $e_h$ is the temporal amplitude envelope (TAE) of the RIR, $a$ is the gain factor, and $\textbf{c}$ is the white Gaussian noise (WGN) carrier. $t_0$ is introduced to promise the causality. Here, $t_0$ is assumed equal to $T_h$.

\begin{figure}[h]
\label{fig:RIRmodels}
 	\centering
\vspace{-0.94em}
 	\centerline{\includegraphics[width=0.42\textwidth]{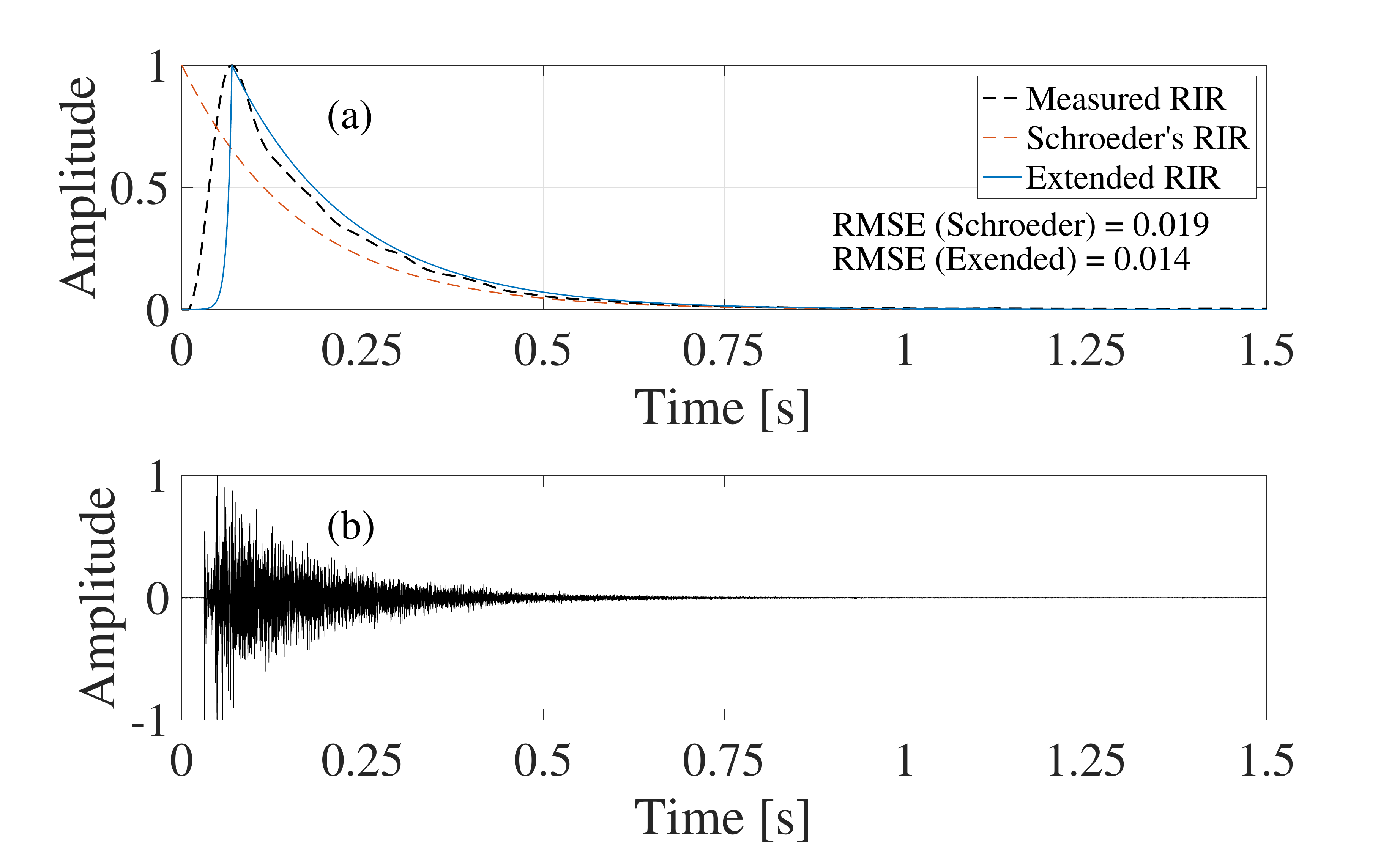}}
 \vspace{-1.4em}
	\caption{Fits of two RIR models with measured RIR: (a) temporal power envelope of RIRs and (b) corresponding RIR.}
\end{figure}

 \vspace{-1.0em}
\subsection{Temporal power envelope model}
\label{ssec:map}
Since we assume a sound field as a linear time-invariant system, the temporal power envelope (TPE) of a reverberation process can be modeled as:
\begin{equation}
	e_y^2(t) = e_x^2(t)*e_h^2(t),
	\label{eq:SystemEnvelope}
\end{equation}
where $e_y^2(t)$ is a TPE of the reverberant signal, $e_x^2(t)$ is a TPE of the input signal, and the asterisk symbol $``*"$ denotes the convolution operation. The TPE of an observed reverberant signal $y(t)$ is extracted as:
\begin{equation}
	e_y^2(t) = \text{LPF}\bigl[\lvert y(t)+j\cdot \text{Hilbert}(y(t))\rvert]^2,
	\label{eq:TPEextraction}
\end{equation}
where \text{LPF} is a low-pass filter. Given the TPE of an input signal based on the superposition principle:
\begin{equation}
	e_x^2(t)= \sum_{k=0}^{K} C_k \cos(2\pi f_{m,k}t+\phi_k), \quad t \in [0,T]
	\label{eq:InputEnvelope}
\end{equation}
where $k$ is the index of $K$ components, $f_{m,k}$ is the modulation frequency at $k$, $\phi_k$ is the phase, $C_k$ is  the constant gain, and $T$ is the time interval. By Eqs.\,(\ref{eq:ext_RIR}) - (\ref{eq:InputEnvelope}), we have the corresponding $e_y^2(t)$ to model the TPE of a reverberant signal. Hence, $e_y^2(t)$ is restored by an inverse-filtering process with a set of $\tilde{T_h}$ and $\tilde{T_t}$. The envelope of the restored TPE $e_{x,\rm{env}}^2(t)$ can be determined as:
\begin{equation}
	 e_{x,\rm{env}}^2(t) = \sum_{k=0}^{K}\frac{C_k T_t}{\tilde{T_t}} \sqrt{\frac{1+\Bigl(\frac{2\pi f_{m,k}\tilde{T_t}}{13.8} \Bigr)}{1+\Bigl(\frac{2\pi f_{m,k}T_t}{13.8}\Bigr)}} \frac{\psi(t,T_h,T_t)}{\psi(t,\tilde{T_h},\tilde{T_t})},
	\label{eq:InverseTPE}
\end{equation}
where $\psi(t,T_h,T_t)\,=\, u(t)-\exp\Bigl[\frac{-13.8(t-T_h)}{T_t}\Bigr] \cos(2\pi f_{m,k}T_h\\+\phi_k)$, and $u(t)$ is the unit-step function. The upper envelope $e_{x,\rm{upr}}^2(t)$ is equal to  $e_{x,\rm{env}}^2(t)$, and the lower envelope  $e_{x,\rm{lwr}}^2(t)$ is equal to $-e_{x,\rm{env}}^2(t)$. 
Eq.\,(\ref{eq:InverseTPE}) indicates that when $T_h=\tilde{T_h}$ and $T_t=\tilde{T_t}$ hold, $e_{x,\rm{upr}}^2(t)\,=\,\sum_{k=0}^{K}C_k$ and $e_{y,\rm{lwr}}^2(t)\,=\,\sum_{k=0}^{K}-C_k$ hold. In this case, the envelopes are irrelevant to time, whereas, when $T_h \neq \tilde{T_h}$ and $T_t \neq \tilde{T_t}$, the envelopes are time-varying. Thus, these time-varying envelopes can be approximated as a first-order polynomial:
\begin{equation}
	e_{x,\rm{upr}}^2(t) = S_{\rm{upr}}t+b_{\rm{upr}}, \ \mbox{and} \quad e_{x,\rm{lwr}}^2(t) = S_{\rm{lwr}}t+b_{\rm{lwr}},
	\label{eq:polynomialEnvleope}
\end{equation}
where $S_{\rm{upr}}$ and $S_{\rm{lwr}}$ are slopes of the envelopes, and $b_{\rm{upr}}$ and $b_{\rm{lwr}}$ are constant factors.
\vspace{-0.9em}
\begin{table}[t]
    \centering
    \vspace{-0.5em}
    \caption{Estimation accuracy of parameters of extended RIR model by proposed and previous methods \cite{Suradej} (RMSE).}
    \begin{tabular}[h]{lcccccc}
    \hline
            & $T_h$  & $T_t$ \\
    \hline
     TAE-CNN & 0.006 & 0.081 \\
     Proposed & 0.087 & 0.193 \\
    \hline
    \end{tabular}
    \label{tabel:ExtResults}
    \vspace{-1.0em}
\end{table}
\vspace{0.5em}
\subsection{Blind estimation method}
\label{ssec:aes}
We estimate the parameters of the RIR by utilizing the model as mentioned in Sec \ref{ssec:map} that generates a TPE of the signals.  From the MTF of the extended RIR model \cite{Suradej}, an infinite-impulse response (IIR) of the inverse filter can be defined as:
\begin{equation}
	E_{h,{\rm{inv}}}(z) = \frac{(1-\alpha z^{-1}) \ (1-\beta z^{-1})}{a^2\ (\alpha-\beta)},
	\label{eq:IIR-MTF}
\end{equation}
where $\alpha=\exp(-13.8/T_t f_s)$, $\beta=\exp(13.8/T_h f_s)$, and $f_s$ is the sampling frequency. 
Whitening, which is key to the AES, is used to transform a complex waveform into a pulse train that consists of even envelopes to use the slopes. The TPE restored at each frame, of which the frame length is $n$, is regarded as autoregressive and rewritten as:
\begin{eqnarray}
	e_x^2[n] &=& -\sum_{i=1}^{p}\sigma_i e_x^2[n-i]+w_x^2[n], \\
	w_x^2[n] &=& \sum_{i=0}^{p}\sigma_ie_x^2[n-i], \ W(z)=\sum_{i=0}^{p}\sigma_iz^{-i},
	\label{eq:autoregressive}
\end{eqnarray}
where $\sigma_i$ is the optimal predictor, $\sigma_0=1$, $p$ is the number of the predictor order, $w_x^2[n]$ is a whitened restored TPE, and $W(z)$ is the frame-based whitening filter. Since Eq.\,(\ref{eq:InverseTPE}) implies that the reverberation smears over all frequencies, where $w_x^2[n] \in e_x^2[n]$, the reverberation smears into $w_x^2[n]$. Therefore, we assert that whitening preserves the reverberation information. 
Hence, the optimal predictor $\sigma_i$ can be determined  by using Wiener-Hopf equations, as used in \cite{Whitening_LPC, sayed2003fundamentals, kailath2000linear}.

The optimal $\widehat{T}_t$ and $\widehat{T}_h$ are specifically obtained. Eq.\,(\ref{eq:AES1}) is derived to determine $\widehat{T}_t$, where ``med" denotes the median operation. Then, substitute $\widehat{T}_t$ into Eq.\,(\ref{eq:IIR-MTF}) to perform inverse filtering so that the $\widehat{T}_h$ can be obtained by using Eq.\,(\ref{eq:AES2}). Table\, \ref{tabel:ExtResults} shows the estimation accuracy of $T_h$ and $T_t$ using the proposed and previous methods \cite{Suradej}. The results show that the proposed method can appropriately estimate the parameters of the extended RIR model.
\begin{eqnarray}
	\widehat{T}_t &=& \underset{T_t}{\text{med}}\bigl\{\underset{T_h,\{\tilde{T_t}\}}{\text{argmin}}\bigl[\log_{10}\bigl(\lvert S_{\rm{upr}} \rvert \bigr)+\log_{10}\bigl(\lvert S_{\rm{lwr}} \rvert \bigr)\bigr]\bigr\}.
	\label{eq:AES1}
	\\
	\widehat{T}_h &=& \text{argmin}\bigl\{\log_{10}\bigl( \lvert S_{\rm{upr}} \rvert \bigr)+\log_{10}\bigl(\lvert S_{\rm{lwr}} \rvert\bigr)\bigr\}.
	\label{eq:AES2}
\end{eqnarray}
\begin{table}[t]
\centering
\vspace{-0.5em}
\caption{Comparison between previous methods \cite{Unoki, Suradej} and proposed method in terms accuracy (RMSE).}
\fontsize{9}{12}\selectfont
\begin{tabular}[h]{lcccccc}
\hline
 & STI & $T_{60}$ & EDT & $C_{80}$ & $D_{50}$ & $T_s$ \cr
\hline
MTF-based & 0.060 & -- & -- & -- & -- & -- \\
TAE-CNN & 0.040 & 0.393 & 0.259 & 2.038 & 12.143 & 0.037\\
Proposed & 0.037 & 0.067 & 0.256 & 2.309 & 14.303 & 0.052\\
\hline
\end{tabular}
\label{tabel:results1}
\vspace{-2.0em}
\end{table}%
\vspace{-1.0em}

Then, we synthesize the estimated RIR\, $\widehat{h}$\,\,by modulating the WGN carrier with the extended RIR model, constructed by that of $\widehat{T}_t$ and $\widehat{T}_h$ using Eq.\,(\ref{eq:ext_RIR}). Finally, the estimated STI and five RAPs are derived from this estimated RIR.
\vspace{-0.5em}
\section{Experiments and results}
\label{sec:expres}
\vspace{-0.5em}
We evaluated the proposed method using reverberant speech signals to confirm whether or not the method can estimate STI and RAPs appropriately. We carried out simulations by using reverberant speech signals synthesized by convolution between speech signals and RIRs from the SMILE dataset, containing $43$ measured RIRs \cite{SMILEdataset}. The speech signals were ten long Japanese sentences uttered by ten speakers (five males and five females) from the ATR dataset \cite{ATRdataset}. We used root-mean-square error (RMSE) as the evaluation metric.

Figure\,\ref{fig:results} shows the results of estimating STI and five-room acoustic parameters from speech signals in realistic reverberant environments. The symbols $``\smallsquare"$, $``\circ"$ and $``\smallstar"$ represent the parameters estimated by the proposed and previous methods, respectively. The horizontal axis indicates the parameters calculated from the RIRs, and the vertical axis indicates the parameters estimated from the speech signals. 

Table\,\ref{tabel:results1} shows the estimation accuracy of the proposed and previous methods. The RMSEs of the estimated STI and $T_{60}$ reveal that the proposed method outperformed the previous methods. With regard to EDT, the estimated results closely approach the ground-truth results calculated from the standard method \cite{ISO3382}. With regard to $C_{80}$, $D_{50}$, and $T_s$, the RMSEs were $2.31$ dB, $14.30$ $\%$, and $0.052$, respectively. However,
it was found that noticeable outliers existed for $C_{80}$, $D_{50}$, and $T_{s}$. These deviations were possibly a result of the carrier signal of the realistic RIRs mismatching the WGN. 
The results demonstrate that the proposed method outperforms or keeps on the same level as the previous works. 
\vspace{-0.5em}
\section{Conclusion}
\label{sec:conclu}
\vspace{-0.5em}
We proposed an analytical method for blindly estimating STI and five RAPs, i.e., $T_{60}$, EDT, $C_{80}$, $D_{50}$, and $T_s$. Instead of relying on training data, we introduced a model to build up a relationship between a reverberant signal and the extended RIR model. This model, based on the TPE of a signal, was used so that the optimal parameters of the RIR model can be estimated. Therefore, the RIR approximated from the RIR model is used to derive STI and RAP. The evaluation results concluded that the proposed method could blindly and simultaneously estimate the STI and RAPs effectively. 
\begin{figure*}[h]
\label{fig:results}
 	\centering
 	\centerline{\includegraphics[width=1.25\textwidth]{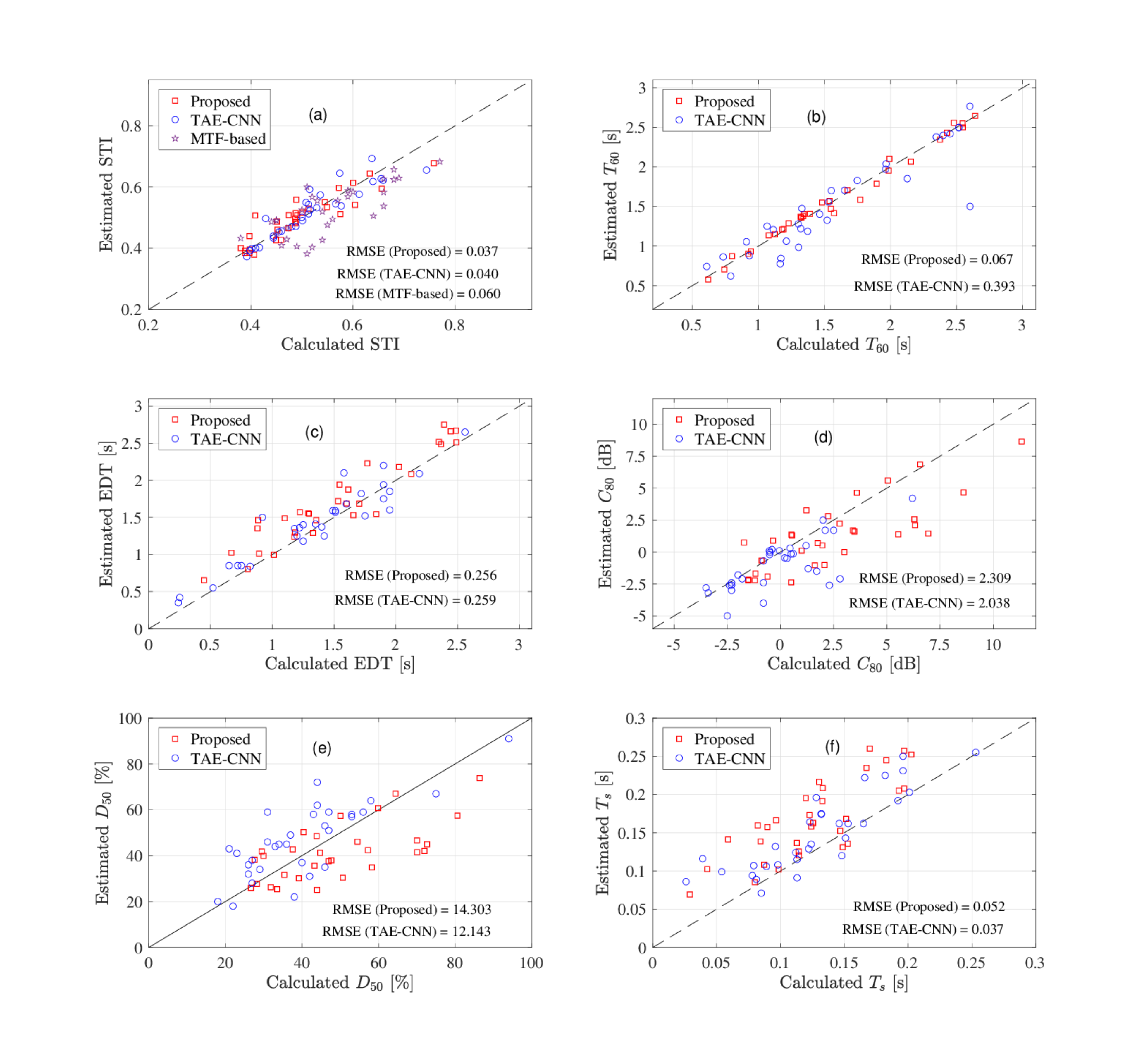}}
 	\vspace{-1.0em}
	\caption{Results of estimating STI and room acoustic parameters from reverberant speech signals: (a) STI, (b) $T_{60}$, (c) EDT, (d) $C_{80}$, (e) $D_{50}$, and (f) $T_s$. $``\smallsquare"$, $``\circ"$, and $``\smallstar"$ denote estimated value of proposed and two previous works, respectively, including TAE-based CNN method (TAE-CNN) \cite{Suradej} and MTF-based method (MTF-based) \cite{Unoki}. Black dashed line represents ground-truth calculated from the RIRs.}
	\label{fig:results}
\end{figure*}

%


\vfill\pagebreak
\bibliographystyle{IEEEbib}
\clearpage
\bibliography{refs}

\end{document}